\documentclass[a4paper,10pt]{article}%
\usepackage{amsmath}
\usepackage{amsfonts}
\usepackage{amssymb}
\usepackage{graphicx}
\usepackage[square,comma,sort&compress]{natbib}%
\providecommand{\U}[1]{\protect\rule{.1in}{.1in}}
\providecommand{\U}[1]{\protect\rule{.1in}{.1in}}
\providecommand{\U}[1]{\protect\rule{.1in}{.1in}}
\providecommand{\U}[1]{\protect\rule{.1in}{.1in}}
\providecommand{\U}[1]{\protect\rule{.1in}{.1in}}
\providecommand{\U}[1]{\protect\rule{.1in}{.1in}}
\providecommand{\U}[1]{\protect\rule{.1in}{.1in}}
\providecommand{\U}[1]{\protect\rule{.1in}{.1in}}
\begin{document}

\title{Bose-Einstein condensates and EPR quantum non-locality}
\author{F.\ Lalo\"{e}\\Laboratoire Kastler Brossel, ENS, CNRS, UPMC;\\24 rue Lhomond, 75005, Paris, France}
\maketitle

\begin{abstract}
The EPR argument points to the existence of additional variables that are
necessary to complete standard quantum theory.\ It was dismissed by\ Bohr
because it attributes physical reality to isolated microscopic systems,
independently of the macroscopic measurement apparatus.\ Here, we transpose
the EPR argument to macroscopic systems, assuming that they are in spatially
extended Fock spin states and subject to spin measurements in remote regions
of space. Bohr's refutation of the EPR argument does not seem to apply in this
case, since the difference of scale between the microscopic measured system
and the macroscopic measuring apparatus can no longer be invoked.

In dilute atomic gases at very low temperatures, Bose-Einstein condensates are
well described by a large population occupying a single-particle state; this
corresponds, in the many particle Hilbert space, to a Fock state (or number
state) with large number $N$. The situations we consider involve two such Fock
states associated to two different internal states of the particles.\ The two
internal states can conveniently be seen as the two eigenstates of the Oz
component of a fictitious spin.\ We assume that the two condensates overlap in
space and that successive measurements are made of the spins of single
particles along arbitrary transverse directions (perpendicular to Oz).

\ In standard quantum mechanics, Fock states have no well defined relative
phase: initially, no transverse spin polarization exists in the system.\ The
theory predicts that it is only under the effect of quantum measurement that
the states acquire a well defined relative phase, giving rise to a transverse
polarization.\ This is similar to an EPR situation with pairs of individual
spins (EPRB), where spins acquire a well defined spin direction under the
effect of measurement - except that here the transverse polarization involves
an arbitrary number of spins and may be macroscopic.\ We discuss some
surprising features of the standard theory of measurement in quantum
mechanics: strong effect of a small system onto an arbitrarily large system
(amplification), spontaneous appearance of a macroscopic angular momentum in a
region of space without interaction (non-locality at a macroscopic scale),
reaction onto the measurement apparatus and angular momentum conservation
(angular momentum version of the EPR argument). \ Bohr's denial of physical
reality for microscopic systems does not apply here, since the measured system
can be arbitrarily large.\ Since here we limit our study to very large number
of particles, no Bell type violation of locality is obtained.

\end{abstract}


\bigskip

The famous Einstein-Podolsky-Rosen (EPR) argument \cite{EPR} considers two
correlated particles, located in two remote regions in space A and B, and
focuses onto the \textquotedblleft elements of reality\textquotedblright%
\ contained in these two regions. It starts from three ingredients: realism,
locality, and the assumption that the predictions of quantum mechanics
concerning measurements are correct\footnote{More precisely, the
EPR\ reasoning only requires that some predictions of quantum mechanics are
correct, those concerning the complete correlations observed between remote
mesurements performed on entangled particles.}; from these inputs it proves
that, to provide a full description of physical reality, standard quantum
mechanics must be completed with additional variables (often called
\textquotedblleft hidden variables\textquotedblright\ for historical reasons).
The EPR argument was refuted by Bohr \cite{Bohr}, who did not accept the
notion of realism introduced by EPR; we give more details on his reply in
\S \ \ref{Bohr}. \ The purpose of the present article is to transpose the
discussion to the macroscopic scale: we investigate situations that are
similar to those considered by EPR but, instead of single particles, we study
Bose-Einstein condensates made of many particles, which can be macroscopic.
For dilute gases, these condensates can be represented as single quantum
states populated with a large, but well defined, number of particles, in other
words by Fock states (number states) with large $N$ (large, but well defined, population).

Several authors \cite{JH, WCW, CGNZ, CD, M-1, M-2, CCT, CH,PS, M3, HB, DRB}
have studied the interference between two such condensates; since the phase of
two Fock states is completely undefined according to standard quantum
mechanics, the question is whether or not a well defined relative phase will
be observed in the interference.\ These authors show that a well defined value
of the relative phase can in fact emerge under the effect of successive
quantum measurements.\ The value taken by this phase is random: it can be
completely different form one realization of the experiment to the next.\ But,
in a given realization, it becomes better and better defined while the
measurements of the position of particles are accumulated. In other words, a
perfectly clear interference pattern emerges from the measurements with a
visible, but completely random and unpredictable, phase.

An interesting variant of this situation occurs if the two highly populated
states correspond to two different internal states of the atoms \cite{SR,
FL-1, MKL}.\ As usual, these two states can conveniently be seen as the two
eigenstates of the $Oz$ component of a fictitious spin.\ One can then study
for instance the situation shown schematically in fig.\ 1, where the two
different internal states with high populations are initially trapped in two
different sites, and then released to let them expand and overlap.\ Many other
situations are also possible; we will discuss some of them in this
article.\ In the overlap region, measurements of the spin component of
particles along directions in the $xOy$ plane are sensitive to the relative
phase of the two condensates.\ A\ free adjustable parameter for every
measurement is the angle $\varphi$ which defines the direction of measurement
in this plane; as discussed in \cite{MKL}, this introduces more flexibility in
choosing a strategy for optimum determination of the phase. Otherwise the
situation is similar to that with spinless particles: initially the relative
relative phase is completely undefined, and nothing can be said about its
value.\ But, as long as the results of the measurements accumulate, the phase
becomes better an better determined under the very effect of the quantum
measurement process.\ This eventually creates a transverse spin polarization
of the whole system, which can be macroscopic for large samples.%
\begin{figure}
[ptb]
\begin{center}
\includegraphics[
height=2.405in,
width=3.4039in
]%
{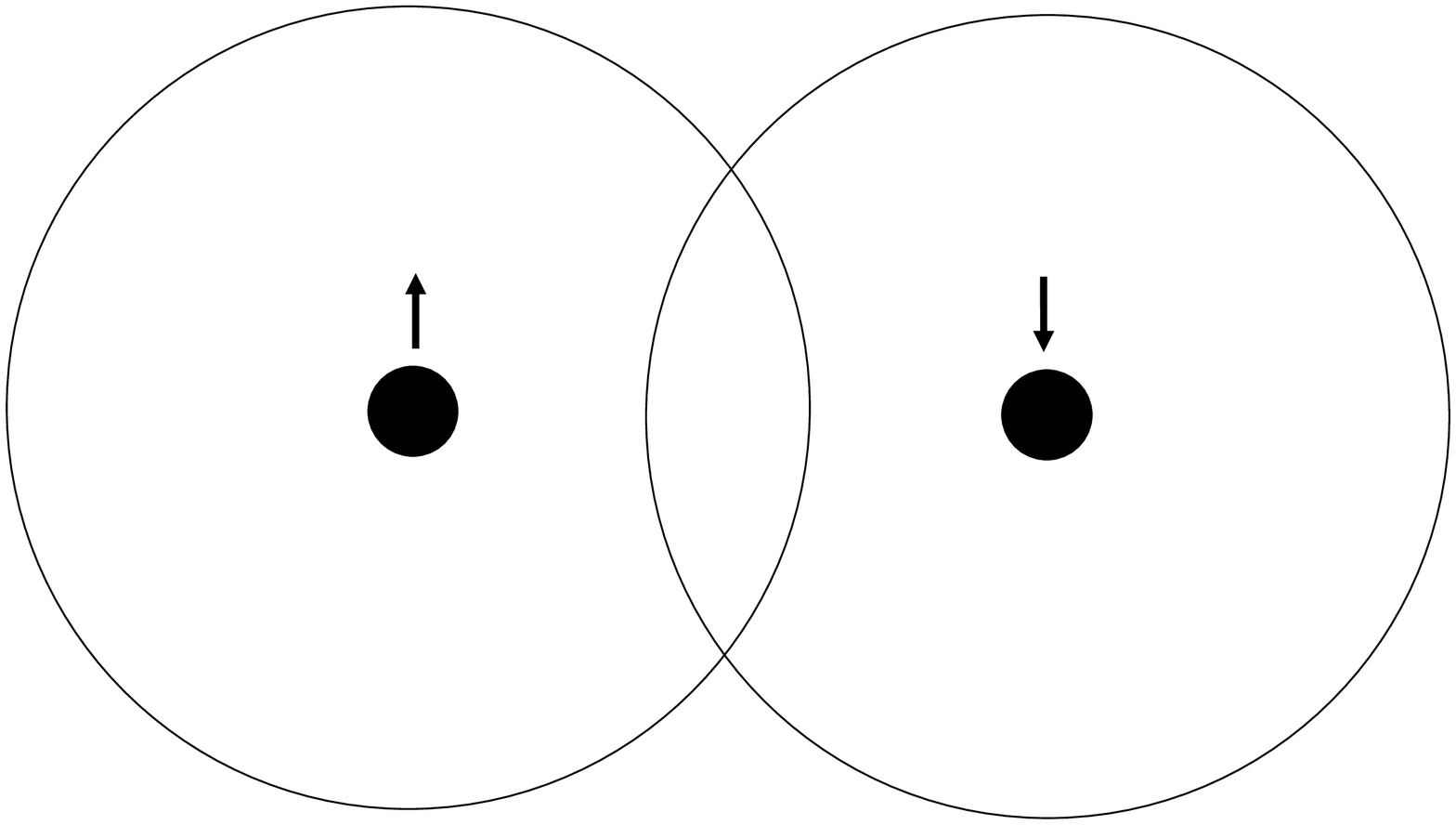}%
\caption{Two condensates, corresponding to two different internal spin states,
expand and overlap in a region of space.\ In this region, particle spin
measurements are performed in transverse directions (perpendicular to the spin
directions of the two initial condensates).}%
\end{center}
\end{figure}

We consider situations where two single particle states associated with
different internal state of atoms overlap in space, and assume that each of
these states has a large population.\ Transverse spin measurements are then
performed in the region of overlap.\ This happens if two Bose-Einstein
condensates are trapped in different sites, and then released to let them expand.

Usually, in the quantum theory of measurement, one emphasizes the role of a
classical macroscopic pointer, the part of the measurement apparatus that
directly provides the information to the human observer.\ Here we have a
curious case where it is the measured system itself that spontaneously creates
a pointer made of a macroscopic number of parallel spins.\ Moreover, for
condensates that are extended in space, we will see that this process can
create instantaneously parallel pointers in remote regions of space, a
situation is obviously reminiscent of the EPR argument in its spin version
given by Bohm (often called EPRB) \cite{B, BA}. We study in this article how
the EPR\ argument can be transposed to this case, and show that the argument
becomes stronger, mostly because the measured systems themselves are now
macroscopic.\ Bohr's refutation, based on the denial of any physical reality
for microscopic systems (cf. \S \ref{Bohr}), then does not apply in the same
way, if it still applies at all.

In \S \ \ref{Bohr} we recall the main features of the EPR argument, which also
gives us the opportunity to summarize Bohr's reply and emphasize his
fundamental distinction between microscopic observed systems and macroscopic
measurement apparatuses.\ In \S \ \ref{calculation}, we introduce the
formalism and generalize the simple calculation of \cite{FL-1}, in particular
to include the case where no particle is detected in the region of
measurement.\ This provides us with the general expression of the joint
probability for any sequence of spin measurements performed in the transverse
direction, and any sequence of results.\ Then \S \ \ref{discussion} contains a
discussion of the physics that is involved: amplification during quantum
measurement, conservation of angular momentum and recoil effects of the
measurements apparatus, quantum non-locality.

\section{EPR argument and its refutation by Bohr\label{Bohr}}

The EPR argument \cite{EPR} \cite{FL-2} considers a physical system made of
two correlated microscopic particles, assuming that they are located in two
remote regions in space A and B where two physicists can perform arbitrary
measurements on them.\ EPR specifically discuss situations where quantum
mechanics predicts that the result of a first measurement performed in A is
completely random, but nevertheless determines with complete certainty the
result of another measurement performed in B.\ They introduce their
\textquotedblleft condition for the reality of a physical
quantity\textquotedblright\ with the famous sentence: \textquotedblleft if,
without in any way disturbing a system, we can predict with certainty the
value of a physical quantity, then there exists an element of physical reality
corresponding to this physical quantity\textquotedblright.\ As a consequence,
just after the measurement in A (but before the measurement in B), since the
result of the second experiment is already certain, an element of reality
corresponding to this certainty must exist in region B.\ But, according to
locality, an element of reality in B can not have been created by the first
measurement performed in region A, at an arbitrarily large distance; the
element of reality necessarily existed even before any measurement.\ Since
standard quantum theory does not contain anything like such a pre-existing
element of reality, it is necessarily an incomplete theory\footnote{Here we
give only the part of the EPR argument that is sometimes called EPR-1: we
consider one type of measurement in each region of space, in other words only
one experimental setup.\ This is sufficient to show that standard quantum
mechanics is incomplete (if one accepts the EPR assumptions). This also
justifies the introduction of statistical averages (or of a variable $\lambda$
that is integrated over initial conditions) in order to prove the Bell
theorem.
\par
In their article, EPR go further and consider several incompatible types of
measurements performed region A.\ They then prove that variables in region B
can have simultaneous realities, even if they are considered as incompatible
in standard quantum mechanics.\ This provides a second proof of
incompleteness, sometimes called EPR-2. Bohr's refutation of the EPR argument
also emphasizes the exclusive character of measurements of incompatible
observables, and therefore concentrates onto EPR-2.
\par
In addition, EPR show in their famous article that, in their views, quantum
mechanics is not only incomplete but also redundant: it can represent the same
physical reality in region B by several differents state vectors (EPR-3).};
the state vector is not sufficient to describe a single realization of an
experiment, but describes only a statistical ensemble of many realizations

Bohr, in his reply \cite{Bohr}, does not criticize the EPR\ reasoning, but the
assumptions on which it is based, which he considers as unphysical.\ He states
that the criterion of physical reality proposed by EPR\ \textquotedblleft
contains an essential ambiguity when applied to quantum
phenomena\textquotedblright\ and that \textquotedblleft their argumentation
does not seem to me\ to adequately meet the actual situation with which we are
faced in atomic physics\textquotedblright\ (here, \textquotedblleft
atomic\textquotedblright\ is presumably equivalent to \textquotedblleft
microscopic\textquotedblright\ in modern language).\ His text has been
discussed by many authors (for an historical review, see for instance
\cite{MJ}), but still remains difficult to grasp in detail (see for instance
Appendix I of \cite{Bell-1}).\ Instead of concentrating his arguments on the
precise situation considered by EPR, Bohr emphasizes in general the
consistency of the mathematical formalism of quantum mechanics and the
\textquotedblleft impossibility of controlling the reaction of the object on
the measuring instruments\textquotedblright.\ But, precisely, the main point
of the EPR argument is to select a situation where these unavoidable
perturbations do not exist! EPR locality implies that a measurement performed
in regions A can create no perturbation on the elements of reality in region B.

Only the second part of Bohr's article really deals with the EPR
argument.\ After stating again that the words \textquotedblleft without in any
way disturbing the system\textquotedblright\ are ambiguous, he concedes that
\textquotedblleft there is of course in a case like that considered (by EPR)
no question of a mechanical disturbance of the system under investigation
during the last critical stage of the measuring procedure\textquotedblright%
.\ Nevertheless, for him what EPR\ have overlooked is that \textquotedblleft
there is essentially the question of an influence on the very conditions which
define the precise types of predictions regarding the future behavior of the
system\textquotedblright\ - the sentence is central but difficult; he probably
means \textquotedblleft an influence \textit{of the measurement performed in A
}on the conditions which define the predictions on the future behavior of the
system \textit{in B}, or maybe \textit{the whole system in both A and
B\textquotedblright}.\ He then states that these conditions are an essential
element of any phenomenon to which the terms \textquotedblleft physical
reality\textquotedblright\ can be attached, and concludes that the EPR proof
of incompleteness is non valid.

J.S.\ Bell summarizes the reply by writing \cite{Bell-1} that, in Bohr's view
\textquotedblleft there is no reality below some classical macroscopic
level\textquotedblright.\ For Bohr, it is incorrect to assign physical reality
to one of the two particles, or even to the group of both particles; physical
reality only has a meaning when macroscopic systems are involved, which here
means the measurement apparatuses.\ He actually attaches physical reality only
to the whole ensemble of the microscopic system and macroscopic measurement
apparatuses, which extends over the two regions A and B of space, and not to
subsystems.\ Then, the EPR\ reasoning, which focuses on B only, becomes
incorrect.\ We remark in passing that Bohr's refutation hinges on the
microscopic character of the measured system, the two quantum particles.

\section{Detecting the transverse direction of spins;
calculation\label{calculation}}

We consider a system composed of particles having two internal states $\alpha$
and $\beta$, which can be seen as the eigenstates of the $Oz$ component of
their spin with eigenvalues $+\hslash$/2 and $-\hslash$/2.\ The particles
populate two quantum states, $\mid u_{a},\alpha>$ (orbital variables described
by an orbital state $\mid u_{a}>$) and $\mid u_{b},\beta>$ (orbital state
$\mid u_{b}>$).\ Initially, the quantum system is in a \textquotedblleft
double Fock state\textquotedblright, with $N_{a}$ particles populating the
first single-particle-state and $N_{b}$ populating the other:
\begin{equation}
\mid\Phi>~=~\left[  \left(  a_{u_{a},\alpha}\right)  ^{\dagger}\right]
^{N_{a}}\left[  \left(  a_{u_{b},\beta}\right)  ^{\dagger}\right]  ^{N_{b}%
}\mid\text{vac}.> \label{1}%
\end{equation}
where $a_{u_{a},\alpha}$ and $a_{u_{b},\beta}$ are the destruction operators
associated with the two single particle states, and $\mid$vac$.>$ is the
vacuum state.\ With the notation of occupation numbers, the same initial state
can also be written:%
\begin{equation}
\mid\Phi>~=~\mid N_{a}:u_{a},\alpha~;~N_{b}:u_{b},\beta~> \label{2}%
\end{equation}
As in \cite{FL-1}\footnote{To correct a sign error in this reference, here we
interchange $\alpha$ and $\beta$.}, we note $\Psi_{\mu}(\mathbf{r})$, with
$\mu=\alpha,\beta$, the field operators associated with internal states
$\alpha$,$\beta$.\ The $\mathbf{r}$ dependent local density operator is then:%
\begin{equation}
n(\mathbf{r})=~~\Psi_{\alpha}^{\dagger}(\mathbf{r})\Psi_{\alpha}%
(\mathbf{r})+\Psi_{\beta}^{\dagger}(\mathbf{r})\Psi_{\beta}(\mathbf{r})
\label{3}%
\end{equation}
while the three components of the local spin density are:%
\begin{equation}%
\begin{array}
[c]{l}%
\sigma_{z}(\mathbf{r})=~\Psi_{\alpha}^{\dagger}(\mathbf{r})\Psi_{\alpha
}(\mathbf{r})-\Psi_{\beta}^{\dagger}(\mathbf{r})\Psi_{\beta}(\mathbf{r})\\
\sigma_{x}(\mathbf{r})=~\Psi_{\alpha}^{\dagger}(\mathbf{r})\Psi_{\beta
}(\mathbf{r})+\Psi_{\beta}^{\dagger}(\mathbf{r})\Psi_{\alpha}(\mathbf{r})\\
\sigma_{y}(\mathbf{r})=~i\left[  \Psi_{\beta}^{\dagger}(\mathbf{r}%
)\Psi_{\alpha}(\mathbf{r})-\Psi_{\alpha}^{\dagger}(\mathbf{r})\Psi_{\beta
}(\mathbf{r})\right]
\end{array}
\label{4}%
\end{equation}
The spin component in the direction of plane $xOy$ making an angle $\varphi$
with $Ox$ is:%
\begin{equation}
\sigma_{\varphi}(\mathbf{r})=e^{-i\varphi}\Psi_{\alpha}^{\dagger}%
(\mathbf{r})\Psi_{\beta}(\mathbf{r})+~e^{i\varphi}\Psi_{\beta}^{\dagger
}(\mathbf{r})\Psi_{\alpha}(\mathbf{r}) \label{5}%
\end{equation}

Suppose now that one measurement is made of the $\varphi$ component of the
spin of particles within a small region of space $\Delta_{\mathbf{r}}$
centered around point $\mathbf{r}$.\ The corresponding operator is:%
\begin{equation}
A(\mathbf{r},\varphi)=~\int_{\Delta_{\mathbf{r}}}d^{3}r^{^{\prime}}%
~\sigma_{\varphi}(\mathbf{r}^{^{\prime}}) \label{6}%
\end{equation}
If the volume $\Delta$ of domain $\Delta_{\mathbf{r}}$ is sufficiently small,
the probability to find more than one particle in this volume is negligible,
and $A(\mathbf{r},\varphi)$ has only three eigenvalues, $0$ and $\pm1$. The
eigenstates corresponding to the eigenvalue $\eta=0$ are all those where
$\Delta_{\mathbf{r}}$ contains no particle; the eigenstates corresponding to
the eigenvalues $\eta=\pm1$ are those for which only one particle is within
$\Delta_{\mathbf{r}}$, in a product state:%
\begin{equation}
\mid\Delta_{\mathbf{r}},\eta>~=~\mid\Delta_{\mathbf{r}}>\otimes\frac{1}%
{\sqrt{2}}\left[  e^{-i\varphi/2}\mid\alpha>+\eta e^{i\varphi/2}\mid
\beta>\right]  \label{7}%
\end{equation}
where $\mid\Delta_{\mathbf{r}}>$ denotes a single particle orbital state with
wave function given by the characteristic function of domain $\Delta
_{\mathbf{r}}$ (equal to $1$ in this domain, $0$ elsewhere).\ In the limit
where the volume $\Delta$ tends to $0$, one can ignore states with more than
one particle in $\Delta_{\mathbf{r}}$, and the $N$ particle states in question
provide a quasi-complete basis. The projector onto eigenvalue $0$ is:%
\begin{equation}
P_{\eta=0}(\mathbf{r})=~1-\int_{\Delta_{\mathbf{r}}}d^{3}r^{^{\prime}%
}~n(\mathbf{r}^{^{\prime}})=\int_{\Delta_{\mathbf{r}}}d^{3}r^{^{\prime}%
}~\left[  \frac{1}{\Delta}-n(\mathbf{r}^{^{\prime}})\right]  \label{8-a}%
\end{equation}
On the other hand, the projectors for eigenvalues $\eta=\pm1$ are:%
\begin{equation}
P_{\eta=\pm1}(\mathbf{r,\varphi})=\,\frac{1}{2}\int_{\Delta_{\mathbf{r}}}%
d^{3}r~\left[  n(\mathbf{r})+\eta\sigma_{\varphi}(\mathbf{r})\right]
\label{8-b}%
\end{equation}

We now consider a series of $K$ measurements, the first of a spin along
direction $\varphi_{1}$ in volume $\Delta_{\mathbf{r}_{1}}$, the second of a
spin along direction $\varphi_{2}$ in volume $\Delta_{\mathbf{r}_{2}}$, etc.,
corresponding to the sequence of operators:%
\begin{equation}
A(\mathbf{r}_{1},\varphi_{1})\text{ \ ; \ }~A(\mathbf{r}_{2},\varphi
_{2})\text{ \ ; \ }~A(\mathbf{r}_{3},\varphi_{3})~;...~;~A(\mathbf{r}%
_{K},\varphi_{K}) \label{9}%
\end{equation}
As in \cite{FL-1}, we assume assume that all $\mathbf{r}$'s are different and
that the regions of measurement $\Delta_{\mathbf{r}_{1}}$, $\Delta
_{\mathbf{r}_{2}}$, ..., $\Delta_{\mathbf{r}_{K}}$ do not overlap, so that all
these operators commute; in addition, and as already mentioned, we assume that
the sequence of measurements is sufficiently brief to ignore any intrinsic
evolution of the system other than the effect of the measurements
themselves.\ Under these conditions, the probability of any sequence of
results:%
\begin{equation}
\eta_{1}=0,\pm1\text{\ \ ; \ }~\eta_{2}=0,\pm1\text{ \ ; \ }...~~~~\eta
_{K}=0,\pm1 \label{10}%
\end{equation}
is simply given by the average value in state $\mid\Phi>$ of the product of
projectors:%
\begin{equation}
<\Phi\mid P_{\eta_{1}}(\mathbf{r}_{1}\mathbf{,\varphi}_{1})\times P_{\eta_{2}%
}(\mathbf{r}_{2}\mathbf{,\varphi}_{2})\times....P_{\eta_{K}}(\mathbf{r}%
_{K}\mathbf{,\varphi}_{K})\mid\Phi> \label{11}%
\end{equation}

When the projectors are replaced by their expressions (\ref{8-a}) and
(\ref{8-b}), with (\ref{5}), we obtain the product of several terms, each
containing various products of field operators.\ In each term, because of the
commutation of the measurements, we can push all $\Psi_{\alpha,\beta}%
^{\dagger}(\mathbf{r})$'s to the left, all $\Psi_{\alpha,\beta}(\mathbf{r})$'s
to the right.\ It is then useful to expand the field operators onto the
annihilation operators for single-particle-states $\mid u_{a},\alpha>$ and
$\mid u_{b},\beta>$:%
\begin{equation}
\Psi_{\alpha}(\mathbf{r})~=~u_{a}(\mathbf{r})\times a_{u_{a},\alpha
}+....~~\text{;}~~~\Psi_{\beta}(\mathbf{r})~=~u_{b}(\mathbf{r})\times
a_{b,\beta}+....~ \label{12}%
\end{equation}
where the terms $+....~$\ symbolize sums over other orbital states that,
together with $u_{a}(\mathbf{r})$, or $u_{b}(\mathbf{r})$, complete a basis in
the orbital space state of a single particle. Since the destruction operators
give zero when they act on states that have zero population, it is easy to see
that all these additional terms simply disappear.\ Each term now contains
between the bra $<\Phi\mid$ and the ket $\mid\Phi>$ a sequence of \ creation
operators, $\left(  a_{u_{a},\alpha}\right)  ^{\dag}$ or $\left(
a_{u_{b},\beta}\right)  ^{\dag}$, followed by another sequence of destruction
operators, $\left(  a_{u_{a},\alpha}\right)  $ or $\left(  a_{u_{b},\beta
}\right)  $. If each state, $u_{a},\alpha$ or $u_{b},\beta$, does not appear
exactly the same number of times in the sequence of creation operators and the
sequence of destruction operators, one obtains the product of two orthogonal
kets, which is zero. If they appear exactly the same number of times, every
creation and destruction operator introduces a factor $\sqrt{\left(
N_{a,b}-q\right)  }$, where $q$ depends on the term considered, but remains
smaller than the number of measurements $K$.\ We assume that:%
\begin{equation}
K\ll N_{a}~,~N_{b} \label{13}%
\end{equation}
which allows us to approximate, as in \cite{FL-1}, all factors $\sqrt{\left(
N_{a,b}-q\right)  }$ by $\sqrt{N_{a,b}}$.

At this point, all operators of $\ \Psi_{\alpha,\beta}^{\dagger}(\mathbf{r})$
are simply replaced by $\sqrt{N_{a,b}}u_{a,b}^{\ast}(\mathbf{r})$, all
$\Psi_{\alpha,\beta}(\mathbf{r})$ by the complex conjugate, but we still have
to take into account the necessity for particle number conservation in each
sequence.\ This can be done by using the mathematical identity:%
\begin{equation}
\int_{0}^{2\pi}\frac{d\Lambda}{2\pi}~e^{in\Lambda}~=~\delta_{n,0} \label{14}%
\end{equation}
(where $n$ is an integer): if we multiply each $\Psi_{\alpha}(\mathbf{r})$ (or
$\sqrt{N_{a}}u_{a}(\mathbf{r})$) by $e^{i\Lambda}$, and each $\Psi_{\alpha
}^{\dagger}(\mathbf{r})$ (or $\sqrt{N_{a}}u_{a}^{\ast}(\mathbf{r})$) by
$e^{-i\Lambda}$, and then integrate $\Lambda$ over $2\pi$, we express the
necessary condition and automatically ensure particle number conservation.

We remark that neither $n(\mathbf{r})$ nor $P_{\eta=0}(\mathbf{r})$ introduce
exponentials of $\Lambda$, since they always contain matched pairs of creation
and destruction operators; exponentials only appear in $\sigma_{\varphi
}(\mathbf{r})$ that is in the projectors $P_{\eta}(\mathbf{r,}$%
\textbf{$\varphi$}$)$ when $\eta=\pm1$. We assume that volume $\Delta$ is
sufficiently small to neglect the variations of the orbital wave functions
over all $\Delta_{\mathbf{r}}$'s. The probability of the sequence of results
(\ref{10}) is then proportional to :%
\begin{equation}%
\begin{array}
[c]{l}%
{\displaystyle\prod\limits_{i}}
\left[  1-\Delta\left(  N_{a}\left\vert u_{a}(\mathbf{r}_{i})\right\vert
^{2}+N_{b}\left\vert u_{b}(\mathbf{r}_{i})\right\vert ^{2}\right)  \right]
\times\\
\times\int_{0}^{2\pi}\frac{d\Lambda}{2\pi}%
{\displaystyle\prod\limits_{j}}
\left\{  \Delta\left[  N_{a}\left\vert u_{a}(\mathbf{r}_{j})\right\vert
^{2}+N_{b}\left\vert u_{b}(\mathbf{r}_{j})\right\vert ^{2}+\eta_{j}\sqrt
{N_{a}N_{b}}\left(  e^{i\left(  \Lambda-\varphi_{j}\right)  }u_{a}%
(\mathbf{r}_{j})u_{b}^{\ast}(\mathbf{r}_{j})+\text{c.c.}\right)  \right]
\right\}
\end{array}
\label{15}%
\end{equation}
where c.c. means complex conjugate. In this expression, the first line
corresponds to the contribution of all results $\eta=0$ (no particle found in
the volumes of detection) and has no $\Lambda$ (or $\varphi$) dependence; the
second line corresponds to all positive detections of spins of particles. It
is convenient to introduce the relative phase of the two wave functions by:%
\begin{equation}
\xi(\mathbf{r})=\arg\left[  u_{a}(\mathbf{r})/u_{b}(\mathbf{r})\right]
\label{16}%
\end{equation}
so that the brackets in the second line become:%
\begin{equation}
\left[  N_{a}\left\vert u_{a}(\mathbf{r}_{j})\right\vert ^{2}+N_{b}\left\vert
u_{b}(\mathbf{r}_{j})\right\vert ^{2}+2\eta_{j}\sqrt{N_{a}N_{b}}_{a}\left\vert
u_{a}(\mathbf{r}_{j})\right\vert \left\vert u_{b}(\mathbf{r}_{j})\right\vert
\cos\left(  \Lambda+\xi(\mathbf{r}_{j})-\varphi_{j}\right)  \right]
\label{17}%
\end{equation}
The contrast of the interference pattern is maximal at points $\mathbf{r}_{j}$
where $\sqrt{N_{a}}\left\vert u_{a}(\mathbf{r}_{j})\right\vert =\sqrt{N_{b}%
}\left\vert u_{b}(\mathbf{r}_{j})\right\vert $, i.e. where the two boson
fields have the same intensity.

These are is the result onto which our discussion below will be based; for a
generalization to spin measurements that are not necessarily in the $xOy$
plane, see the appendix of \cite{MKL}.

\section{Physical discussion\label{discussion}}

\subsection{Role of the $\Lambda$ integral}

Suppose that we consider a sequence where only one spin is detected; the
product over $j$ in (\ref{15}) then contains only one bracket, summed over
$\Lambda$ between $0$ and $2\pi$.\ The contribution of each value of $\Lambda$
gives nothing but the probability of the two results, $\eta_{j}=\pm1$, for a
spin $1/2$ that is described by a density matrix $\rho$ given by:%
\begin{equation}
\rho\propto\left(
\begin{array}
[c]{ll}%
N_{a}\left\vert u_{a}(\mathbf{r}_{1})\right\vert ^{2} & \sqrt{N_{a}N_{b}}%
u_{a}^{\ast}(\mathbf{r}_{j})u_{b}(\mathbf{r}_{1})e^{-i\Lambda}\\
\sqrt{N_{a}N_{b}}u_{a}(\mathbf{r}_{1})u_{b}^{\ast}(\mathbf{r}_{1})e^{i\Lambda}
& N_{b}\left\vert u_{b}(\mathbf{r}_{1})\right\vert ^{2}%
\end{array}
\right)  \label{18}%
\end{equation}
This is easily checked by calculating the trace of the product of $\rho$ by
the projector:%
\begin{equation}
\left[  1+\eta\left(  e^{-i\varphi}\sigma_{+}+e^{i\varphi}\sigma_{-}\right)
\right]  /2 \label{19}%
\end{equation}
The $Oz$ component of the spin before measurement is then proportional to
$N_{a}\left\vert u_{a}(\mathbf{r}_{1})\right\vert ^{2}-N_{b}\left\vert
u_{b}(\mathbf{r}_{1})\right\vert ^{2}$, its transverse component proportional
to $2\sqrt{N_{a}N_{b}}_{a}\left\vert u_{a}(\mathbf{r}_{1})\right\vert
\left\vert u_{b}(\mathbf{r}_{1})\right\vert $, with an azimuthal direction
specified by angle $\Lambda-\xi(\mathbf{r}_{1})$. Now, since $\Lambda$ is
summed between $0$ and $2\pi$, the off diagonal elements disappear from
(\ref{18}), meaning that this azimuthal direction is initially completely
random; the spin loses its transverse orientation and keeps only its $Oz$
component.\ This is natural since we are starting from Fock states with
completely undetermined relative phase. Therefore, for the first transverse
measurement the two results $\pm1$ always have the same probability, and the
adjustable parameter $\varphi_{1}$ plays no role.

Now consider a sequence with two measurements and two results $\eta_{1,2}%
=\pm1$.\ In (\ref{15}), the $\Lambda$ integral then introduces
correlations.\ The result of the first measurement provides an information on
the probabilities of the results of the second: this information is contained
in a $\Lambda$ distribution that is given by (\ref{17}), with $\eta_{j}$
replaced by the result $\pm1$ of the first measurement, and $\mathbf{r}_{j}$
replaced by the point $\mathbf{r}_{1}$ at which this measurement was made. The
information is still not very precise, since the width of the $\Lambda$
distribution is of the order of $\pi$; but, for instance, if the first result
was $+1$ and if the two angles of measurements $\varphi_{1}$ and $\varphi_{2}$
are close or even equal, there is more chance to find again $+1$ than $-1$ for
the result of the second measurement.

When more and more spin measurements are obtained, the $\Lambda$ distribution
becomes narrower and narrower, meaning that more and more information on the
value of $\Lambda$ is accumulated.\ Standard quantum mechanics considers that
$\Lambda$ has no physical existence at the beginning of the series of
measurements, and that its determination is just the result of a series of
random perturbations of the system introduced by the
measurements.\ Nevertheless, (\ref{15}) shows that all observations are
totally compatible with the idea of a pre-existing value of $\Lambda$, which
is perfectly well defined but unknown, remains constant, and is only revealed
(instead of being created) by the measurements. For a more detailed discussion
of the evolution of the $\Lambda$ distribution, and of the optimum strategy
concerning the choice of the angles of measurement $\varphi_{j}$ to better
determine $\Lambda$, see ref.\ \cite{MKL}.

It is interesting to find a situation where an additional (hidden) variable
$\Lambda$ emerges so naturally from a standard calculation in quantum
mechanics.\ It appears mathematically as a way to express the conservation of
number of particles.\ In other words, the role of the additional variable is,
by integration, to ensure the conservation of the conjugate variable. This
contrasts with usual theories with additional variables, where they are
introduced more or less arbitrarily, the only constraint being that the
statistical average over the new variables reproduces the predictions of
standard quantum mechanics.

\subsection{Small and big condensates; amplification during measurement}

In our calculation, we have made no special assumption concerning the orbital
wave functions $u_{a}(\mathbf{r})$ and $u_{b}(\mathbf{r})$ associated with the
two highly populated single particle quantum states; they can overlap much or
little in space, and in many ways.\ We will consider situations where their
configuration leads to the discussion of interesting physical effects.\ For
simplicity, from now on we assume that the two wave functions have the same
phase at every point of space $\mathbf{r}$, so that $\xi(\mathbf{r})$
vanishes.\ This simplification occurs if the two states correspond to
stationary states trapped in a real potential, as often the case in
experiments with Bose-Einstein condensates; it is convenient, but not
essential\footnote{For instance, we exclude the case where two condensates are
still expanding, as in figure 1 and as in the experiment described in ref.
\cite{WMEH}.\ Much of what we write can nevertheless be transposed to such
cases, in terms of the phase of an helicoidal structure of the spin directions
in space, instead of just parallel spins.}.%
\begin{figure}
[ptb]
\begin{center}
\includegraphics[
height=2.1309in,
width=3.0139in
]%
{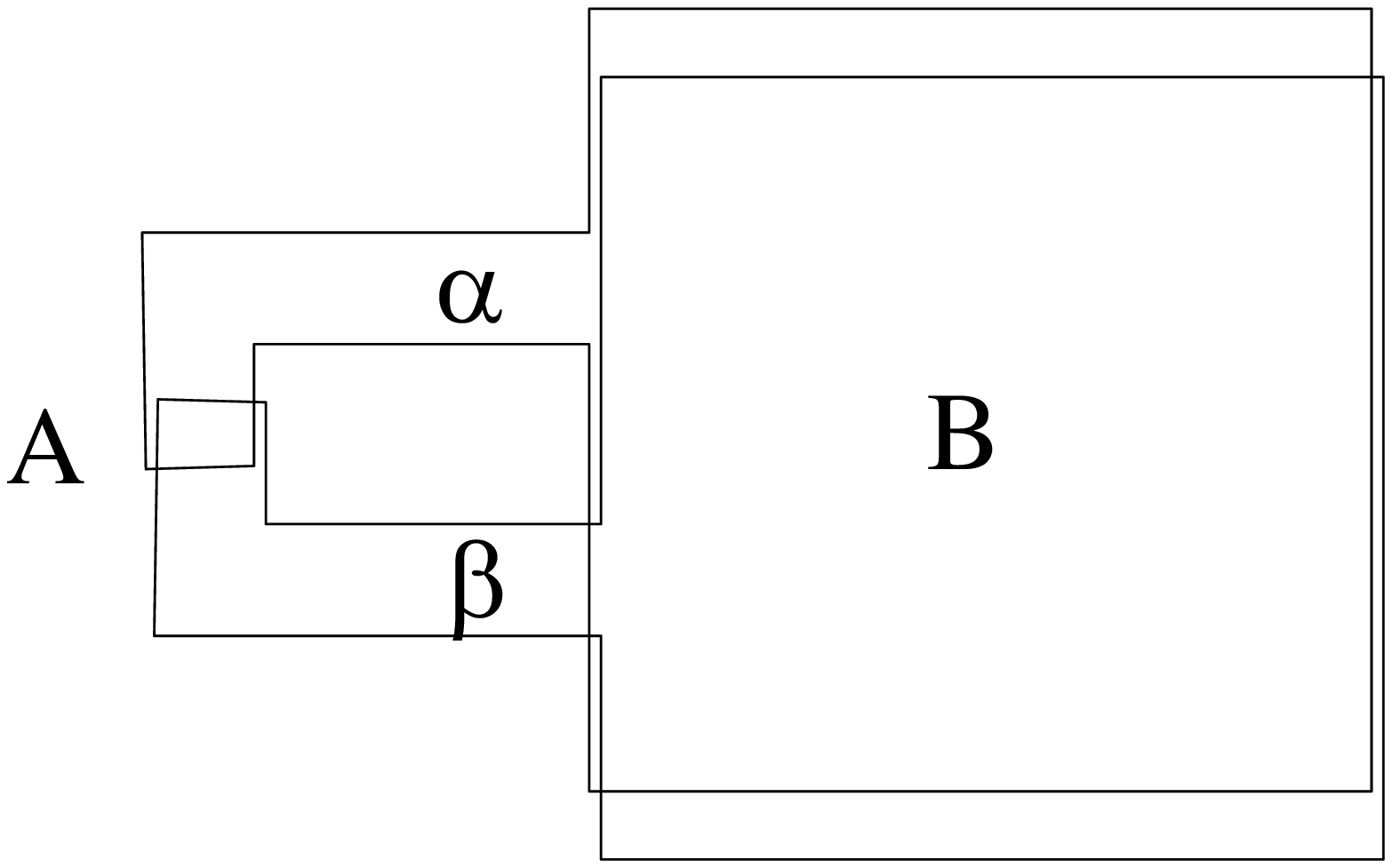}%
\caption{Two different spin states ($\alpha=+$ and $\beta=-$) are associated
with two orbital wave functions that overlap mostly in a large region B, but
also have \textquotedblleft fingers\textquotedblright\ that overlap in a much
smaller region A.\ The two states are macroscopically populated.\ Under the
influence of a few measurements of the spin of particles performed in region
A, a macroscopic transverse spin polarization appears in region B.}%
\end{center}
\end{figure}

As a first example, we consider two states such as those represented
schematically in figure 2.\ The states are mostly located in a region of space
B where they strongly overlap, but also have \textquotedblleft
fingers\textquotedblright\ that overlap in another small region of space A,
where the spin measurements are actually performed.\ We assume that A is not
too small, and contains an average number of spins (100 for instance) that
remains sufficient to perform several measurements and determine the relative
phase of the two Fock states with reasonable accuracy. On the other hand, the
number of spins in region B may be arbitrarily large, of the order of the
Avogadro number for instance.

In this situation, our preceding calculation applies and predicts that the
measurement of the spins in A will immediately create a spontaneous
polarization in B that is parallel to the random polarization obtained in
A.\ In other words, standard quantum mechanics predicts a giant amplification
effect, where the measurement performed on a few microscopic particles induces
a transverse polarization in a macroscopic assembly of spins. In itself, the
idea is not too surprising, even in classical mechanics: one could see the
assembly of spins in B as a metastable system, ready to be sensitive to the
tiny perturbation of a microscopic system in A.\ In this perspective, the
perturbations created by the measurement in A would propagate towards B and
trigger its evolution towards a given spin direction. But this is not the
context in which we have obtained the prediction: we have not assumed any
evolution of the state vector of the system between one measurement and the
next.\ In fact, what standard quantum mechanics describes here is not
something that propagates along the state and has a physical mechanism (such
as, for instance, the propagation of Bogolubov phonons in the condensates); it
is just \textquotedblleft something with no time duration\textquotedblright%
\ that is a mere consequence of the postulate of quantum measurement (wave
packet reduction).

Leggett and Sols \cite{LS, L} discuss a similar situation in the context of
two large superconductors, which acquire a spontaneous phase by the creation
of a Josephson current between them, which in turn is measured by a tiny
compass needle in order to obtain its phase.\ Here again we have a small
system determining the state of a much larger system, without any physical
mechanism.\ These authors comment the situation in the following terms:
\textquotedblleft can it really be that by placing, let us say, a minuscule
compass needle next to the system, with a weak light beam to read off its
position, we can force the system to realize a definite macroscopic value of
the current?\ Common sense rebels against this conclusion, and we believe that
in this case common sense is right\textquotedblright. They then proceed to
explain that the problem may arise because we are trying to apply to
macroscopic objects quantum postulates that were designed 80 years ago for the
measurements of microscopic objects, because other measurements were not
conceivable then. In other words, we are trying to use present standard
quantum mechanics beyond its range of validity.\ They conclude that what is
needed in a new quantum measurement theory.

What is interesting to note, as we have already mentioned in the introduction,
is that here we have a case where the measured system itself creates a
macroscopic pointer, made of a large assembly of parallel spins, that directly
\textquotedblleft shows\textquotedblright\ the direction of the spins
resulting from the measurements.\ Usually, in the theory of quantum
measurement, this pointer is the last part of the measuring apparatus, not
something that interacts directly with the measured system itself, or even
less is part of it.

\subsection{EPR\ non locality with Fock states}

Now suppose that the two condensates have the shape sketched in figure 3,
extending over a large distances, and overlapping only in two remote regions
of space A and B. Again, the number of particles in both regions is arbitrary,
and in particular can be macroscopic in B.\ We have a situation that is
similar to the usual EPR\ situation: measurements performed in A can determine
the direction of spins in both regions A and B.\ If we rephrase the
EPR\ argument to adapt it to this case, we just have to replace the words
\textquotedblleft before the measurement in A\textquotedblright\ by
\textquotedblleft before the series of measurements in A\textquotedblright,
but all the rest of the reasoning remains exactly the same: since the elements
of reality in B can not appear under the effect of what is done at an
arbitrary distance in region A, these elements of reality must exist even
before the measurements performed in A.\ Since the initial double Fock state
of quantum mechanics does not contain any information on the direction of
spins in B, this theory is incomplete.%
\begin{figure}
[ptb]
\begin{center}
\includegraphics[
height=2.2338in,
width=3.16in
]%
{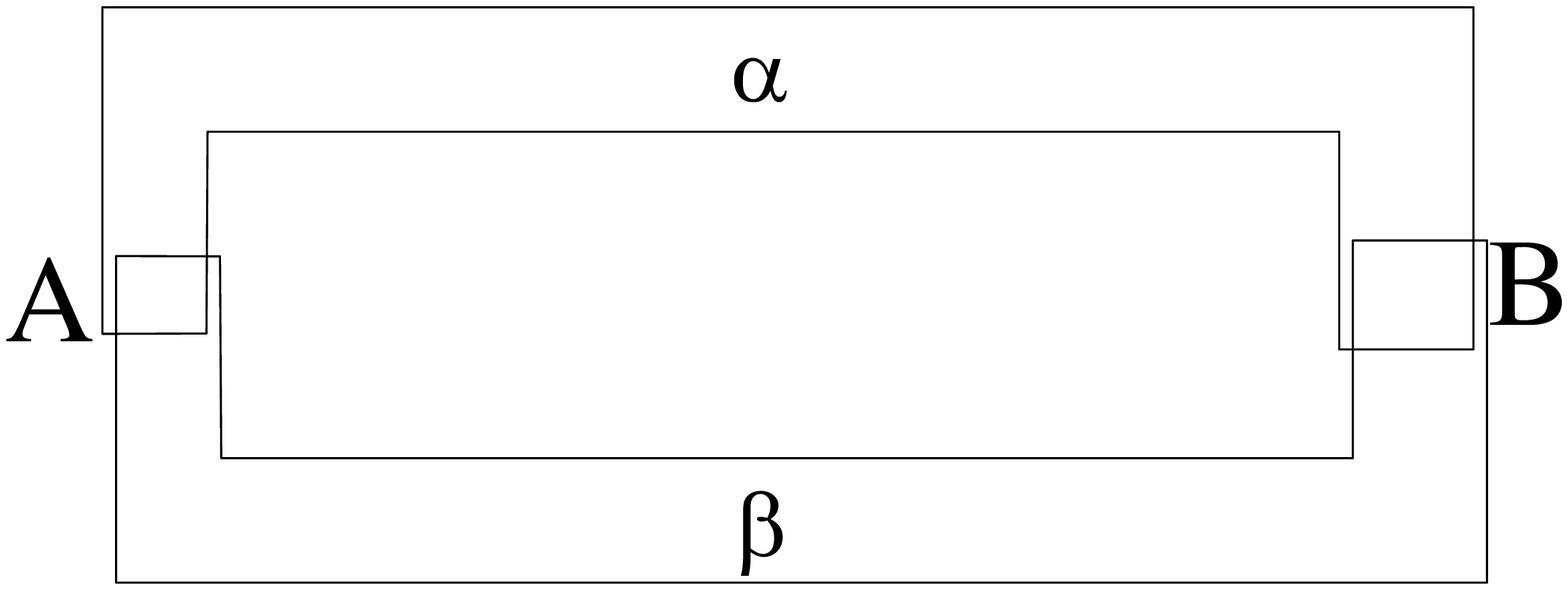}%
\caption{Two highly populated Fock states associated with opposite spin
direction ($\alpha=+$ and $\beta=-$) overlap in two remote regions A\ and B of
space.\ A series of transverse spin measurements in A triggers the appearance
of a well defined transverse orientation in A, and also that of a parallel
macroscopic transverse orientation in B (quantum non locality). This
corresponds to an angular momentum that seems to appear in B from nothing,
with no interaction at all.}%
\end{center}
\end{figure}

What is new here is that the EPR elements of reality in B correspond to a
system that is macroscopic.\ One can no longer invoke its microscopic
character to deprive the system contained in B of any physical reality! The
system can even be at our scale, correspond to a macroscopic magnetization
that can be directly observable with a hand compass; is it then still possible
to state that it has no intrinsic physical reality?\ When the EPR argument is
transposed to the macroscopic world, it is clear that Bohr's refutation does
no longer apply in the form written in his article; it has to be at least
modified in some way.

Another curiosity, in standard quantum mechanics, is that it predicts the
appearance of a macroscopic angular momentum in region B without any
interaction.\ This seems to violate angular momentum conservation.\ Where does
this momentum come from? Usually, when a spin is measured and found in some
state, one considers that the angular momentum is taken as a recoil by the
measurement apparatus.\ When the measured system is microscopic and the
apparatus macroscopic, the transfer of angular momentum is totally negligible
for the latter, so that there is no hope to check this idea; but, at least,
one can use the idea as a theoretical possibility.\ Here, the situation is
more delicate: what is the origin of the angular momentum that appears in B
during measurement?\ Could it be that the apparatus in A, because the system
in A is entangled with a macroscopic system in B, takes a macroscopic recoil,
even if it measures a few spins only?\ A little analysis shows that this is
impossible without introducing the possibility for superluminal communication:
the recoil in A would allow to obtain information on B (if the states have
been dephased locally for instance).\ So, it can not be the measurement
apparatus in A that takes the angular momentum recoil corresponding to
B.\ Then, if we believe that angular momentum can not appear in a region of
space without interactions, even during operations that are considered as
\textquotedblleft measurements\textquotedblright\ in standard quantum
mechanics, this leads us to an "angular momentum EPR proof": we are forced to
conclude that the transverse polarization of the spins in B already existed
before any measurement started\footnote{To avoid this conclusion, one can
either give up angular momentum conservation in measurements (making them even
more special physical processes than usually thought!), or take the Everett
interpretation (\textquotedblleft relative state\textquotedblright\ or
\textquotedblleft many minds\textquotedblright\ interpretation) where no
transverse polarization ever appears in B, even after the measurements..}%
.\ Since this is not contained in the double Fock state, standard quantum
mechanics is incomplete.
\begin{figure}
[ptb]
\begin{center}
\includegraphics[
height=2.5676in,
width=3.6322in
]%
{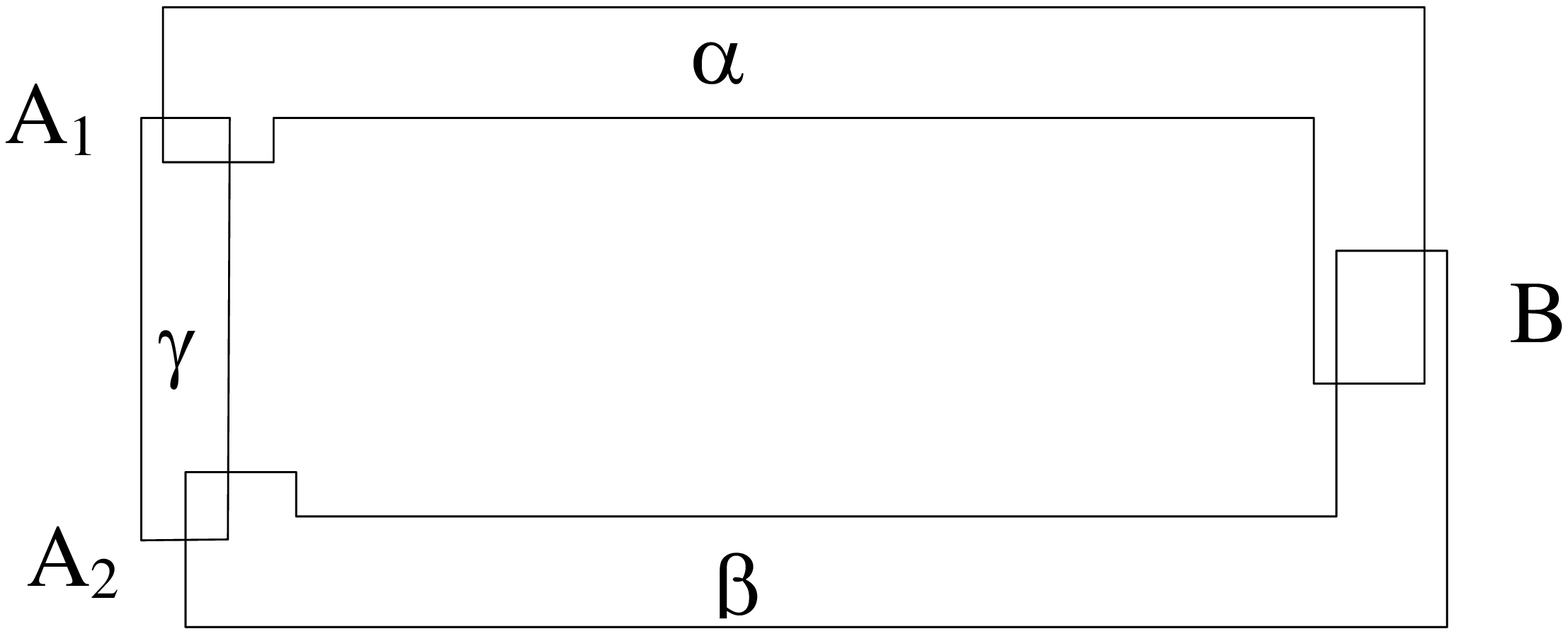}%
\caption{In a variant of the situation shown in Fig.\ 3, measurements are
performed in region A, at points A$_{1}$ and A$_{2}$, of the relative phase
between a small local condensate (in internal state $\gamma$) and two extended
condensates (respectively in internal states $\alpha$ and $\beta$).\ By
transitivity, this determines the relative phase of the large condensates,
resulting by interference in a transverse spin polarization in region B.\ Even
if the measurements with the local A condensate do not involve angular
momentum (assumed to have no matrix elements between the overlapping states),
this operation can create a macroscopic angular momentum in region B.}%
\end{center}
\end{figure}

We can make the argument even more convincing by using the scheme sketched in
figure 4. We now assume that the two condensates in internal states $\alpha$
and $\beta$ overlap in B but not in A, where they both overlap both with the
same third condensate in a third internal state $\gamma$. We furthermore
assume that angular momentum has matrix elements between $\alpha$ and $\beta$,
but not between $a$ and $\gamma$ and between $\beta$ and $\gamma$ (for
instance, the parity of $\gamma$ may be the opposite of that of the two other
internal states): the transverse measurements in A correspond to some
observable that has appropriate matrix elements and parity, electric dipole
for instance. We know (see for instance \cite{FL-1}) that phase determination
in Fock states is transitive: fixing the phase between $a$ and $\gamma$ on the
one hand, between $\beta$ and $\gamma$ on the other, will determine the
relative phase of $\alpha$ and $\beta$. Under these conditions, in standard
quantum mechanics, the macroscopic angular momentum that appears in B can be a
consequence of measurements in A of physical quantities that have nothing to
to with angular momentum, so that the measurement apparatuses have no reason
to take any angular momentum recoil at all.\ Still, they create a large
angular momentum in B.\ Again, if we do not accept the idea of angular
momentum appearing from nothing, we must follow EPR and accept that the
angular momentum was there from the beginning, even if we had no way to
predict its direction\footnote{As above, the only other logical possibility is
to choose the other extreme: the Everett interpretation, where no angular
momentum exists even after the measurements.}.

\section{Possible objections}

In this text, we have discussed thought experiments, not attempted to propose
feasible experiments.\ We have just assumed that the states that are necessary
for the discussion can be produced, and that they are sufficiently robust to
undergo a series of measurements, with no other perturbation than the
measurements themselves; this may require that the sequence of measurements be
sufficiently fast.\ Of course, one could always object that these double Fock
states are fundamentally not physical, for instance because some selection
rule forbids them.\ This would be in contradiction with the generally accepted
postulate that all quantum states belonging to the space of states (Hilbert
space) of any physical system are accessible.\ If this postulate is true,
there should be no fundamental reason preventing the preparation of a double
Fock state, even for a system containing many particles.

A second objection could be that these states may exist but be so fragile
that, in practice, it will always be impossible to do experiments with them.
In the context of second order phase transitions and spontaneous symmetry
breaking, Anderson \cite{A-1, A-2, A-3} has introduced the notion of
spontaneous phase symmetry breaking for superfluid Helium 4 and
superconductors.\ According to this idea, coupled superfluid systems at
thermal equilibrium are not in Fock states: as soon as they become superfluid
by crossing the second order transition, some unavoidable small perturbation
always manages to transform the simple juxtaposition of the two Fock states
into a single coherent Fock state containing all the $N_{a}+N_{b}$ particles,
for which the two quantum states have a well defined relative phase.\ This
assumption is for example implicit in the work of Siggia and Ruckenstein
\cite{SR}, where the two condensates are considered as having a well defined
phase from the beginning\footnote{It is interesting to note in passing that,
unexpectedly, Anderson's spontaneous symmetry breaking concept is so closely
related to the old idea of hidden/additional variables in quantum
mechanics.\ A specificity, nevertheless, is that Anderson sees the additional
variables as appearing during second order superfluid phase transitions.}.

In our discussion, we have assumed neither the existence of a second order
phase transition nor even thermal equilibrium, just the availability of the
large initial double Fock state. Is there any general mechanism that favours
coherent states over double Fock states? Decoherence may actually inroduce
this preference.\ As in ref.\ \cite{PS}, we can introduce the so called
coherent \textquotedblleft phase states\textquotedblright\ by:%
\begin{equation}
\mid\phi,N>~=\frac{1}{\left(  2^{N}N!\right)  ^{1/2}}~\left[  e^{i\phi
/2}\left(  a_{u_{a},\alpha}\right)  ^{\dagger}+e^{-i\phi/2}\left(
a_{u_{b},\beta}\right)  ^{\dagger}\right]  ^{N}\mid\text{vac}.> \label{20}%
\end{equation}
in terms of which the ket $\mid\Phi>$ of (\ref{2}) can be written:%
\begin{equation}
\mid N_{a}:u_{a},\alpha~;~N_{b}:u_{b},\beta~>\propto\int_{0}^{2\pi}%
d\phi~e^{i\left(  N_{b}-N_{a}\right)  \phi}\mid\phi,N> \label{21}%
\end{equation}
If $M$ is large, the phase state (\ref{20}) have a macroscopic transverse
orientation in an azimuthal direction defined by $\phi$; this orientation is
likely to couple to the external environment, as most macroscopic variable
do.\ For instance, if the spin of particles is associated with a magnetic
moment, the different phase states create different macroscopic magnetic
fields that will affect at least some microscopic particles of the
environment, transferring them into states that are practically orthogonal for
different values of $\phi$. In other words, the basis of phase states is the
\textquotedblleft preferred basis\textquotedblright\ for the system coupled to
its environment.\ As a consequence, the coherent superposition (\ref{21})
spontaneously transforms into a superposition where each component, defined by
a very small $\phi$ domain, is correlated with a different state of the
environment.\ The correlation quickly propagates further and further into the
environment, without any limit as long as the Schr\"{o}dinger equation is
obeyed (this is the famous Schr\"{o}dinger cat paradox).\ As a result, the
observation of interference effects between different $\phi$ values becomes
more and more difficult, in practice impossible.\ In terms of the the trace of
the density operator over the environment, the coherent superposition
(\ref{21}) decays rapidly into an incoherent mixture of different $\phi$
states. For a general discussion of the observability of macroscopically
distinct quantum states, see for instance ref. \cite{AJL}.

Decoherence is unavoidable, but does not really affect our conclusions.\ It
just means that, in the standard interpretation, when the measurements are
performed in region A and determine the transverse polarization, they fix at
the same time the spin directions in B as well as the state of the local
environment.\ The real issue is not coherence, or the coupling to the
environment; it is the emergence of a single macroscopic result, which is
considered as an objective fact and a result of the observation in the
standard interpretation (but of course not in the Everett interpretation). In
the end, decoherence is not an essential issue in our discussion.

\ A third objection might be size limitations: are there inherent limits to
the size of highly populated Fock states and Bose-Einstein condensates?\ Is
there any reason why large sizes should make them extremely sensitive to small
perturbations?\ One could think for instance of thermal fluctuations that may
introduce phase fluctuations and put some temperature dependent limit on the
size of the coherent system.\ Other possible mechanisms, such as
inhomogeneities of external potentials, might break the condensate into
several independent condensates, etc. Generally speaking, we know that ideal
condensed gases are extremely sensitive to small perturbations\footnote{For
instance, condensate in ideal gases tend to localize themselves in tiny
regions of space \cite{LN}.\ Nevertheless, this is a pathology introduced by
the infinite compressibility of the condensate in an ideal gas; it disappears
as soon as the atoms have some mutual repulsion.}, but fortunately also that
repulsive interactions between the atoms tend to stabilize condensed systems.
They do not only introduce a finite compressibility of the condensate, but
also tend to stabilize the macroscopic occupation of a single quantum state
\cite{PN}.\ This should increase the robustness of large systems occupying a
unique single Fock state, even if extended in space.

Experimentally, Bose-Einstein condensates in dilute gases at very low
temperatures provide systems that are very close to being in a highly
populated Fock state.\ Nevertheless, until now experiments have been performed
with gas samples that are about the size of a tenth of a millimeter; one can
therefore not exclude that new phenomena and unexpected perturbations will
appear when much larger condensates are created. In any case, even if the
non-local effects that we have discussed are limited to a range of a tenth of
a millimeter (or any other macroscopic length), they remain non-local effects
on which a perfectly valid EPR\ type argument can be built!

\section{Conclusion}

We can summarize the essence of this article by saying that, in some quantum
situations where macroscopic systems populate Fock states with well defined
populations, the EPR argument becomes significantly stronger than in the
historical example with two microscopic particles.\ The argument speaks
eloquently if favour of a pre-existing relative phase of the two states -
alternatively, if one prefers, of an interpretation where the phase remains
completely undetermined even after the measurements (Everett interpretation) -
but certainly not in favour of the orthodox point of view where the phase
appears during the measurements. If we stick to this orthodox view, surprising
non-local effects appear in the macroscopic world.\ These effects can be
expressed in various ways, including considerations on macroscopic angular
momentum conservation, but not in terms of violations of Bell type
inequalities (this is because the form of the $\Lambda$ integral in
(\ref{15}), with positive terms inside it, automatically ensures that the Bell
inequalities are satisfied).\ In any case, Bohr's denial of physical reality
of the measured system alone becomes much more difficult to accept when this
system is macroscopic.\ Of course, no one can predict what Bohr would have
replied to an argument involving macroscopic spin assemblies, and whether or
not he would have maintained his position concerning the emergence of a single
macroscopic result during the interaction of the measured system with the
measurement apparatuses.

Another conclusion is that quantum mechanics is indeed incomplete, not
necessarily in the exact sense meant by EPR, but in terms of the postulates
related to the measurement: they do not really specify what is the reaction of
the measured system on the measurement apparatus (\textquotedblleft
recoil\ effect\textquotedblright).\ Ignoring this reaction was of course
completely natural at the time when quantum mechanics was invented: only
quantum measurements of microscopic systems were conceivable at that time, so
that these effects were totally negligible.\ But now this is no longer true,
so that we need a more complete theory for quantum measurement on a
macroscopic system \textquotedblleft in which all the assumptions about
relative energy and time scales, etc.. are made explicit and if necessary
revised\textquotedblright\ \cite{LS}. Bose-Eintein condensates in gases seem
to be good candidates to explore this question theoretically and experimentally.

\bigskip

\begin{center}
\textbf{Acknowledgments}
\end{center}

The author is grateful to W.\ Mullin, A.\ Leggett, C.\ Cohen-Tannoudji and
J.\ Dalibard for useful discussions and comments.

\begin{center}

\end{center}

\end{document}